\documentclass[prl,aps,twocolumn,showpacs,superscriptaddress]{revtex4-1}
\usepackage{amssymb}
\usepackage{amsmath}
\usepackage[version=4]{mhchem}
\usepackage{hyperref}
\hypersetup{colorlinks=true, citecolor=blue, filecolor= black, linkcolor= blue, urlcolor= blue}
\usepackage{graphicx}
\usepackage{dcolumn}
\usepackage{bm}
\usepackage{float}
\usepackage{braket}
\begin{document}

\title{Helium Induced Nitrogen Salt at High Pressure}

\author{Jingyu Hou}
\affiliation{Center for High Pressure Science, State Key Laboratory of Metastable Materials Science and Technology, School of Science, Yanshan University, Qinhuangdao 066004, China}
\affiliation{Key Laboratory of Weak-Light Nonlinear Photonics and School of Physics, Nankai University, Tianjin 300071, China}

\author{Xiao-Ji Weng}
\affiliation{Key Laboratory of Weak-Light Nonlinear Photonics and School of Physics, Nankai University, Tianjin 300071, China}

\author{Artem R. Oganov}
\affiliation{Skolkovo Institute of Science and Technology, Skolkovo Innovation Center, 3 Nobel Street, Moscow 143026, Russia}

\author{Xi Shao}
\affiliation{Center for High Pressure Science, State Key Laboratory of Metastable Materials Science and Technology, School of Science, Yanshan University, Qinhuangdao 066004, China}

\author{Guoying Gao}
\affiliation{Center for High Pressure Science, State Key Laboratory of Metastable Materials Science and Technology, School of Science, Yanshan University, Qinhuangdao 066004, China}

\author{Xiao Dong}
\email{xiao.dong@nankai.edu.cn}
\affiliation{Key Laboratory of Weak-Light Nonlinear Photonics and School of Physics, Nankai University, Tianjin 300071, China}

\author{Hui-Tian Wang}
\affiliation{Key Laboratory of Weak-Light Nonlinear Photonics and School of Physics, Nankai University, Tianjin 300071, China}

\author{Yongjun Tian}
\affiliation{Center for High Pressure Science, State Key Laboratory of Metastable Materials Science and Technology, School of Science, Yanshan University, Qinhuangdao 066004, China}

\author{Xiang-Feng Zhou}
\email{xfzhou@nankai.edu.cn}
\email{zxf888@163.com}
\affiliation{Center for High Pressure Science, State Key Laboratory of Metastable Materials Science and Technology, School of Science, Yanshan University, Qinhuangdao 066004, China}
\affiliation{Key Laboratory of Weak-Light Nonlinear Photonics and School of Physics, Nankai University, Tianjin 300071, China}


\begin{abstract}
\noindent
The energy landscape of helium-nitrogen mixtures is explored by \textit{ab initio} evolutionary searches, which predicted several stable helium-nitrogen compounds in the pressure range from 25 to 100 GPa. In particular, the monoclinic structure of HeN$_{22}$ consists of neutral He atoms, partially ionic dimers N$_{2}$$^{\delta-}$, and lantern-like cages N$_{20}$$^{\delta+}$. The presence of helium not only greatly enhances structural diversity of nitrogen solids, but also tremendously lowers the formation pressure of nitrogen salt. The unique nitrogen framework of (HeN$_{20}$)$^{\delta+}$N$_{2}$$^{\delta-}$ may be quenchable to ambient pressure even after removing helium. The estimated energy density of N$_{20}$$^{\delta+}$N$_{2}$$^{\delta-}$ (10.44 kJ/g) is $\sim$2.4 times larger than that of trinitrotoluene (TNT), indicating a very promising high-energy-density material.
\end{abstract}



\maketitle
Helium is the most inert and second most abundant element in the universe. Possessing the highest among all elements ionization potential (25 eV) and zero electron affinity, both of which are related to its closed-shell electronic structure, it generally does not interact with other elements or compounds under ambient conditions \cite{R01}. However, chemistry of the elements can be greatly altered by pressure \cite{R02,R03,R04,R05,R06}. Even an element as inert as helium can display some reactivity at high pressure. For example, the already synthesized Na$_{2}$He, as well as the predicted Na$_{2}$HeO, FeHe$_{x}$, and other helium-bearing compounds possess unusual properties, some of them being electrides, some $йд-$ superionic conductors, and such compounds may trap helium in the interiors of the Earth and other planets \cite{R06,R07,R08,R09,R10,R11,R12}. Hence the high-pressure helium-bearing materials are of great interest to fundamental physics, chemistry, Earth and space sciences. At the same time, helium reacts with elements and compounds and forms solid van der Waals (vdW) compounds such as He(N$_{2}$)$_{11}$ \cite{R13}, NeHe$_{2}$ \cite{R14}, and He@H$_{2}$O \cite{R12,R15}. For these materials, the enthalpy of formation is close to zero and the removal of helium has little effect on the host's properties. At high pressure, helium occupies empty spaces in the structure, helping to improve its packing efficiency. This should stabilize those crystal structures that have cavities of suitable size to host helium, and can be used to synthesize such structures at lower pressures than in the absence of helium.

At ambient conditions, pure nitrogen exists in the form of a gas of diatomic molecules. The nature of nitrogen is dramatically altered under high pressure because the particularly strong triple $\ce{N#N}$ bond transforms into a much weaker single N-N bond. The average bond energy of $\ce{N#N}$ (945 kJ/mol) is almost 6 times larger than single N-N (160 kJ/mol), thus the reversion of single N-N to triple $\ce{N#N}$ would be accompanied with a huge energy release. Consequently, singly-bonded nitrogen-rich compounds are excellent candidates for high-energy-density materials. The search for polymeric nitrogen or nitrogen-rich materials has attracted enormous attention due to the potential applications for energy storage, explosives, and propellants \cite{R16,R17,R18,R19,R20,R21,R22,R23,R24,R25,R26,R27,R28}. Direct synthesis of single-bonded polymeric nitrogen is extremely challenging. For instance, the successfully synthesized polymeric nitrogen allotrope has the so called cubic gauche structure (denoted as $cg$-N); it was synthesized in 2004 \cite{R16}, long after its prediction, and is metastable at least down to 25 GPa \cite{R19,R20}. After that, layered polymeric nitrogen with $Pba2$ symmetry (denoted as $LP$-N) and $P4_{2}bc$ symmetry (denoted as $HLP$-N) were synthesized in succession \cite{R25,R28}. Overall, the calculated high-pressure zero-temperature phase diagram of polymeric nitrogen was extensively studied with the established sequence of $cg$-N $\rightarrow$ $LP$-N (188 GPa) $\rightarrow$ N$_{10}$ (263 GPa) $\rightarrow$ $Cmca$ metal (2.1 TPa) $\rightarrow$ $P4/nbm$ metallic salt (2.5 TPa) \cite{R28}. However, the challenge remains to preserve polymeric nitrogen at ambient conditions. Therefore, helium-nitrogen mixtures may provide an alternative way to achieve this goal at high pressure \cite{R07,R13,R29,R30}. Solid vdW compound He(N$_{2}$)$_{11}$ was successfully synthesized at room temperature and persisted up to 135 GPa \cite{R13,R29,R30}, but we are interested primarily in the polymeric compounds of helium and nitrogen, and presumed that they may appear at mild pressures and high temperatures.

\begin{figure}[t]
\begin{center}
\includegraphics[width=7.0cm]{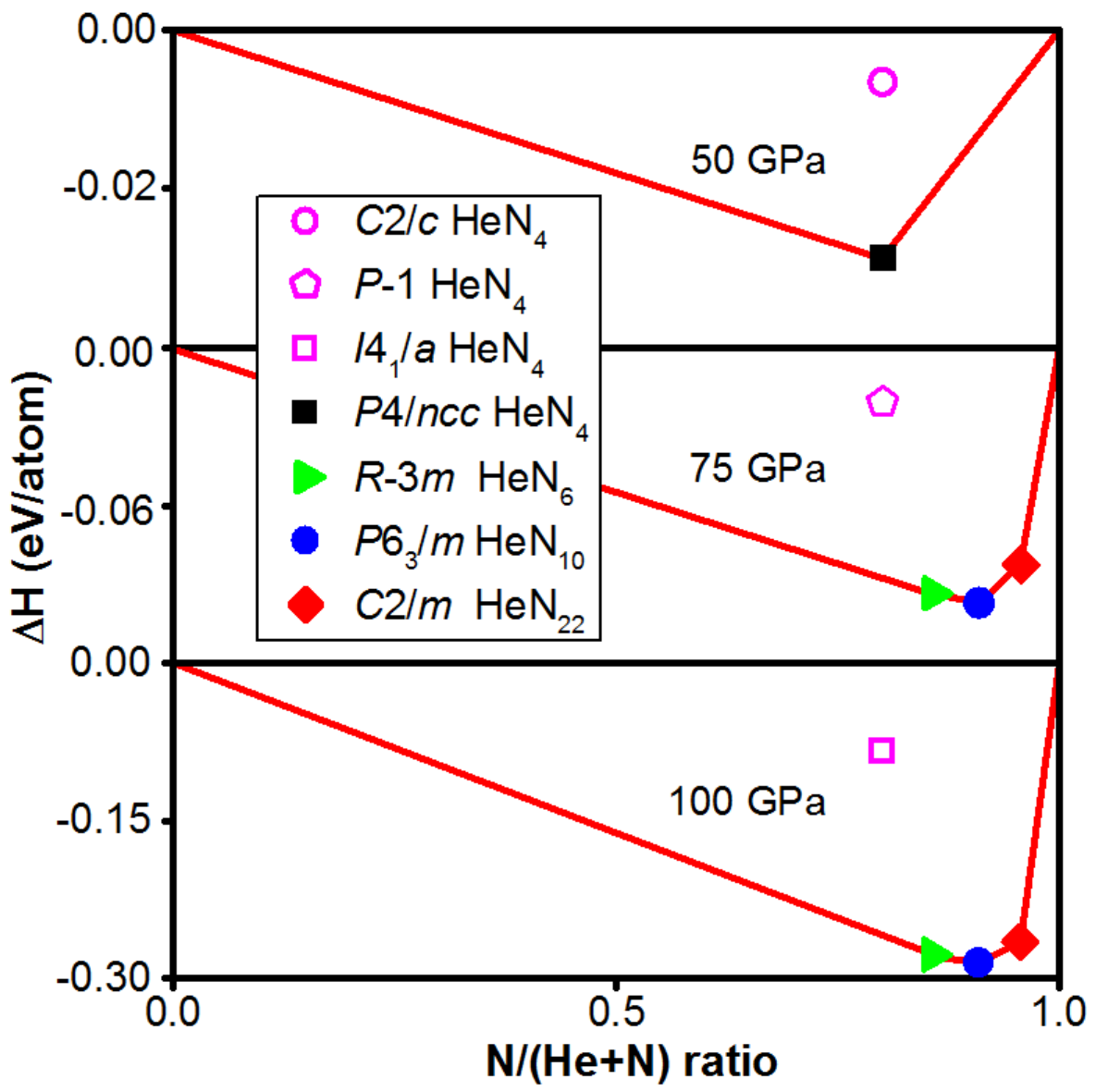}
\caption{\label{Figure1}
 Predicted convex hulls of the He-N system at selected pressures, using $hcp$-He and $\epsilon$-N$_{2}$ structures for reference. The reported structures of HeN$_{4}$ \cite{R07} with symmetries $C2/c$, $P\overline{1}$ and $I4_{1}/a$ are plotted for comparison.}
\end{center}
\end{figure}

The variable-composition evolutionary algorithm \textsc{uspex} \cite{R31,R32} was utilized to predict all stable He-N compounds. Our searches had up to 40 atoms per primitive cell and were performed at pressures of 25, 50, 75, and 100 GPa. For all stable compounds (i.e. those located on the convex hull), fixed-composition evolutionary searches were carried out to ensure the most stable structure. A material is deemed stable at zero temperature if its enthalpy of formation ($\Delta$H) from either elements or any isochemical mixture of other compounds is negative \cite{R02,R03,R06,R33,R34}. Here the formation enthalpy is defined as $\Delta$H = H(He$_{1-x}$N$_{x}$)$-$(1-$x$)H(He)$-$$x$H(N). Note that the calculated transition pressure from molecular $\epsilon$-N$_{2}$ to solid $cg$-N is about 56 GPa [Fig.~S1(a)] \cite{R35}. However, due to the large activation barrier, $\epsilon$-N$_{2}$ experimentally transformed into $cg$-N at pressures above 110 GPa and temperatures above 2000 $K$ \cite{R19}, and thus the structures of hexagonal close packing helium ($hcp$-He) and $\epsilon$-N$_{2}$ are used as pure elements \cite{R07}. Structure relaxations and calculations of the electronic properties were conducted in the framework of density functional theory (DFT) within the generalized gradient approximation using Perdew-Burke-Ernzerhof functional \cite{R36}, as implemented in the \textsc{vasp} code \cite{R37}. We employed the all-electron projector augmented wave \cite{R38} pseudopotentials with valence shells 1$s^{2}$ and 2$s^{2}$2$p^{3}$ for He and N, respectively. The plane-wave cutoff energy of 500 eV and uniform $\Gamma$-centered $k$ meshes with the resolution of $2 \pi \times 0.06$~\AA$^{-1}$ were adopted for \textsc{uspex} searches. They are further increased to 700 eV and $2 \pi \times 0.03$~\AA$^{-1}$ for precise energy and property calculations for stable structures. To examine the importance of vdW interactions on energetic stability, the semiempirical dispersion-correction method \cite{R39} was used but produced practically identical results. Phonon dispersion curves were calculated with the \textsc{phonopy} code \cite{R40} using the finite displacement method. In addition, a crystal orbital Hamiltonian population (COHP) method \cite{R41} implemented in the \textsc{lobster} package \cite{R42} was utilized for chemical bonding analysis (including the Mulliken and L\"{o}wdin population analysis).

The convex hulls in Fig.~\ref{Figure1} show several stable helium-nitrogen compounds at different pressures. First, a vdW compound HeN$_{4}$ with $P4/ncc$ symmetry is stable in the pressure range from 0 to 68 GPa (denoted as $P4/ncc$ HeN$_{4}$ and similar designation was applied for other compounds. i.e., $P6_{3}/m$ HeN$_{10}$, $C2/m$ HeN$_{22}$, and $R\overline{3}m$ HeN$_{6}$). Second, other unprecedented polymeric forms, namely, $P6_{3}/m$ HeN$_{10}$, $C2/m$ HeN$_{22}$, and $R\overline{3}m$ HeN$_{6}$, are stabilized above 64 GPa. The pressure-composition phase diagram of the He-N system is shown in Fig.~S1(b) \cite{R35}. The corresponding stabilization pressures for $P6_{3}/m$ HeN$_{10}$, $C2/m$ HeN$_{22}$, and $R\overline{3}m$ HeN$_{6}$ are 64.3, 67.6, and 74 GPa, respectively. The only synthesized He-N compound has the stoichiometry He(N$_{2}$)$_{11}$ and its structure had been determined by single-crystal X-ray diffraction ($P6_{3}/m$ HeN$_{22}$) \cite{R30}. However, we are unable to study such structure within the present \textit{ab initio} calculations because of the orientationally disordered N$_{2}$ molecules and its fractional atomic occupations. Nevertheless, the successful synthesis of $P6_{3}/m$ HeN$_{22}$ suggests it as a suitable precursor for synthesis of of $C2/m$ HeN$_{22}$ or other similar polymorphs.

\begin{figure}[t]
\begin{center}
\includegraphics[width=7.5cm]{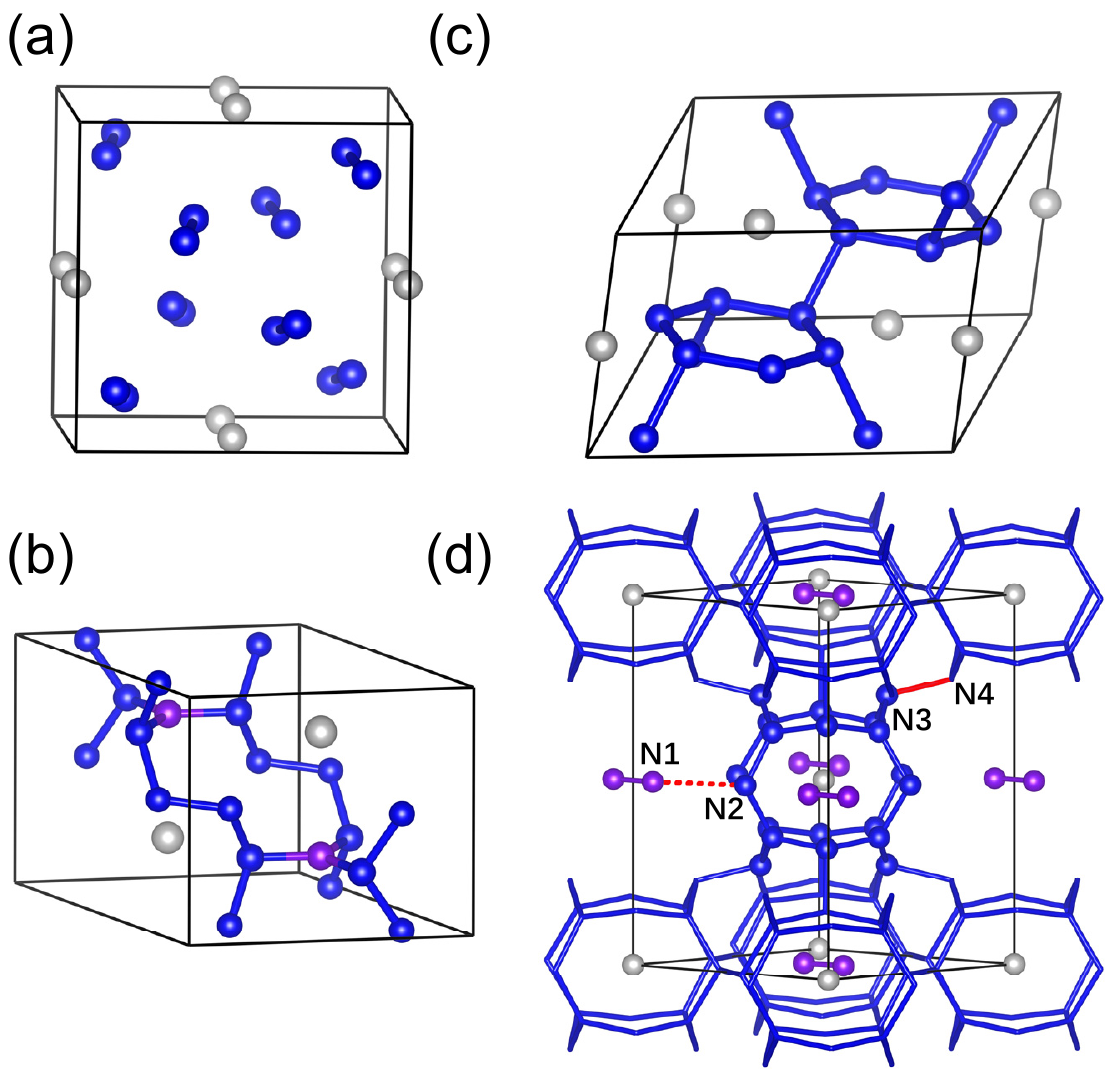}
\caption{\label{Figure2}
 Crystal structures of (a) $P4/ncc$ HeN$_{4}$ at 50 GPa, (b) $P6_{3}/m$ HeN$_{10}$ at 100 GPa, (c) $R\overline{3}m$ HeN$_{6}$ at 100 GPa, and (d) $C2/m$ HeN$_{22}$ at 100 GPa. To illustrate the structural feature more clearly, the lattice constants of $C2/m$ HeN$_{22}$ are redefined \cite{R35}. The He atoms, N$_{2}$ dimers in HeN$_{22}$ and $sp^{2}$-N in HeN$_{10}$ are colored in gray and purple, respectively. The nearest distance between N$_{2}$ dimers and N$_{20}$ cages was represented by N1-N2 (red dotted line), while the shortest inter-cage N$_{20}$ bond was represented by N3-N4.}
\end{center}
\end{figure}

Lattice parameters and atomic positions of various structures are listed in Table SI \cite{R35}. As shown in Fig.~\ref{Figure2}(a), $P4/ncc$ HeN$_{4}$ is a tetragonal vdW compound with eight N$_{2}$ molecules per unit cell toward different orientations. The length of triple $\ce{N#N}$ bonds in $P4/ncc$ HeN$_{4}$ is $\sim$1.1~{\AA} at 50 GPa, the same as in $\epsilon$-N$_{2}$ and $C2/c$ HeN$_{4}$. The different packing of N$_{2}$ molecules in $P4/ncc$ HeN$_{4}$ results in greater density, smaller volume and lower enthalpy of formation than $C2/c$ HeN$_{4}$ at all pressures, explaining why the previously predicted $C2/c$ HeN$_{4}$ is not stable \cite{R07}. Band structure from GGA-PBE calculation shows that $P4/ncc$ HeN$_{4}$ is an insulator with band gap of 4.92 eV at 50 GPa [Fig.~S2(a)] \cite{R35}. As pressure increases above 64.3 GPa, the triple $\ce{N#N}$ bonds break and transform into single N-N bonds. The polymeric structure of $P6_{3}/m$ HeN$_{10}$ emerges and is followed by $C2/m$ HeN$_{22}$ and $R\overline{3}m$ HeN$_{6}$. They may coexist at 75 GPa and above. As shown in Fig.~\ref{Figure2}(b), $P6_{3}/m$ HeN$_{10}$ has three inequivalent nitrogen atoms with Wyckoff positions $2d$, $6h$ and $12i$, the former having $sp^{2}$ hybridization, while the other sites are $sp^{3}$. The $sp^{2}$-N has planar triangular coordination shape, whereas the $sp^{3}$-N is pyramidal in shape. There is a distinct difference in the distribution of lone pairs between $sp^{2}$-N and $sp^{3}$-N. Electronic localization function (ELF) shows that lone pairs symmetrically locate at both sides of $sp^{2}$-N while gather one side on top of $sp^{3}$-N (Fig.~S3) \cite{R35}. Such coexistence of $sp^{2}$-N and $sp^{3}$-N was reported in diamondoid N$_{10}$ structure above 263 GPa \cite{R23}. In contrast, the presence of He causes similar behavior in $P6_{3}/m$ HeN$_{10}$ at much lower pressure ($\sim$64 GPa). Band structure shows that $P6_{3}/m$ HeN$_{10}$ is an insulator with band gap of 4.05 eV at 100 GPa [Fig.~S2(b)] \cite{R35}.

$R\overline{3}m$ HeN$_{6}$ is stable in the pressure range from 74 GPa to at least 110 GPa, and its structure consists of He and $sp^{3}$-N simultaneously. The nitrogen framework is made up of the distorted N$_{6}$ hexagons [Fig.~\ref{Figure2}(c)]. Each N atom has threefold coordination: two bonds are used to form N$_{6}$ hexagon and have length of 1.31~{\AA} at 100 GPa, while the other connects N$_{6}$ hexagons and has length of 1.41~{\AA}, which corresponds to a slightly compressed single N-N bond. Since the intrahexagonal N-N bonds are stronger (shorter) than those interhexagonal bonds, such peculiar bonding configurations indicate that neutral aromatic hexazine N$_{6}$ may be observed on decompression, if the weak interhexagonal N-N bonds break first. Band structure shows that $R\overline{3}m$ HeN$_{6}$ is an insulator with band gap of 3.42 eV at 100 GPa [Fig.~S2(c)] \cite{R35}.

Since the stoichiometry of HeN$_{22}$ had been determined experimentally, $C2/m$ HeN$_{22}$ has attract more interest among four newly predicted phases. As shown in Fig.~\ref{Figure2}(d), $C2/m$ HeN$_{22}$ is composed of encapsulated He, interstitial N$_{2}$ dimers, and polymeric lantern-like N$_{20}$ cages. The bond length in the N$_{2}$ dimer is 1.09~{\AA}, the strongest bonds in this compound. The exotic lantern-like N$_{20}$ cage, it contains twenty N atoms and is made of six-connected irregular N$_{8}$ rings with D$_{2h}$ symmetry [Fig.~\ref{Figure2}(d)]. The intra-cage bond lengths have a range from 1.28 to 1.34~{\AA} at 100 GPa, while the inter-cage bond lengths have a range from 1.38 to 1.42~{\AA}, indicating that inter-cage N-N bonds are weaker than intra-cage bonds. One helium atom is enclosed in the center of each N$_{20}$ cage (the nearest
He-N is $\sim$1.96~{\AA}). The nearest distance between N$_{2}$ dimer and N$_{20}$ cage is $\sim$2.2~{\AA}; no covalent bond exists between these two units, but there is an electrostatic interaction.

Based on the above mentioned unique structures, Bader charge analysis \cite{R43} was performed to investigate the charge transfer. Bader charges for He and N in $P4/ncc$ HeN$_{4}$ at 50 GPa, $R\overline{3}m$ HeN$_{6}$ and $P6_{3}/m$ HeN$_{10}$ at 100 GPa are between 0 and $|0.05~e|$. In the same way, Bader charge for He in $C2/m$ HeN$_{22}$ is about $-$0.05~$|e|$ at 100 GPa and keeps almost unchanged at various pressures, implying no real charge transfer between He and N. Unexpectedly, there is a considerable charge transfer between N$_{2}$ dimers and N$_{20}$ cages at 100 GPa. As summarized in Table SII \cite{R35}, the N$_{2}$ dimers are negatively charged ($-$0.19 ~$|e|$) and serve as anions, while the positively charged N$_{20}$ cages (+0.24~$|e|$) act as cations. Since helium is enclosed in the center of N$_{20}$ cage, the structure of $C2/m$ HeN$_{22}$ could be regarded as (HeN$_{20}$)$^{\delta+}$N$_{2}$$^{\delta-}$ ($\delta$ $\approx$ 0.24~$|e|$ at 100 GPa). The value of $\delta$ becomes large on increasing pressures ($\delta$ $\approx$ 0.27~$|e|$ at 150 GPa), while it is about 0.15~$|e|$ at zero pressure. To ensure the magnitude of charge transfer between N$_{2}$ dimers and N$_{20}$ cages, the calculated Mulliken charge and L\"{o}wdin charge \cite{R42} are $\sim$0.22~$|e|$ and 0.25~$|e|$, respectively. All results from different methods are consistent with each other. Previously, all-nitrogen ``metallic salt" N$_{2}$$^{\delta+}$N$_{5}$$^{\delta-}$ ($\delta$ = 0.37~$|e|$) was predicted stable above 2.5 TPa with $P4/nbm$ symmetry \cite{R24}. Crystal structure of (HeN$_{20}$)$^{\delta+}$N$_{2}$$^{\delta-}$ has the charge transfer comparable to that of N$_{2}$$^{\delta+}$N$_{5}$$^{\delta-}$ (2.5 TPa) but at much lower pressure (68 GPa). We therefore suggest that the framework of N$_{20}$$^{\delta+}$N$_{2}$$^{\delta-}$ resembles a new kind of elemental salt $йд-$ the nitrogen clathrate salt. Emergence of charge transfer had also been observed in $\gamma$ boron \cite{R44}, yet this is a different case from nitrogen. Boron is electron-deficient and forms icosahedron-based structures. The multicenter bonding and the charge transfer on some B-B bonds play a decisive role in structural diversity and stability. In contrast, nitrogen is an electron-rich element. The packing efficiency of lone pair electrons is crucial for the formation of various high-pressure phases rather than charge transfer, e.g., $cg$-N, diamondoid structure of N$_{10}$ and so on \cite{R19,R23}.

\begin{figure}[t]
\begin{center}
\includegraphics[width=8.0cm]{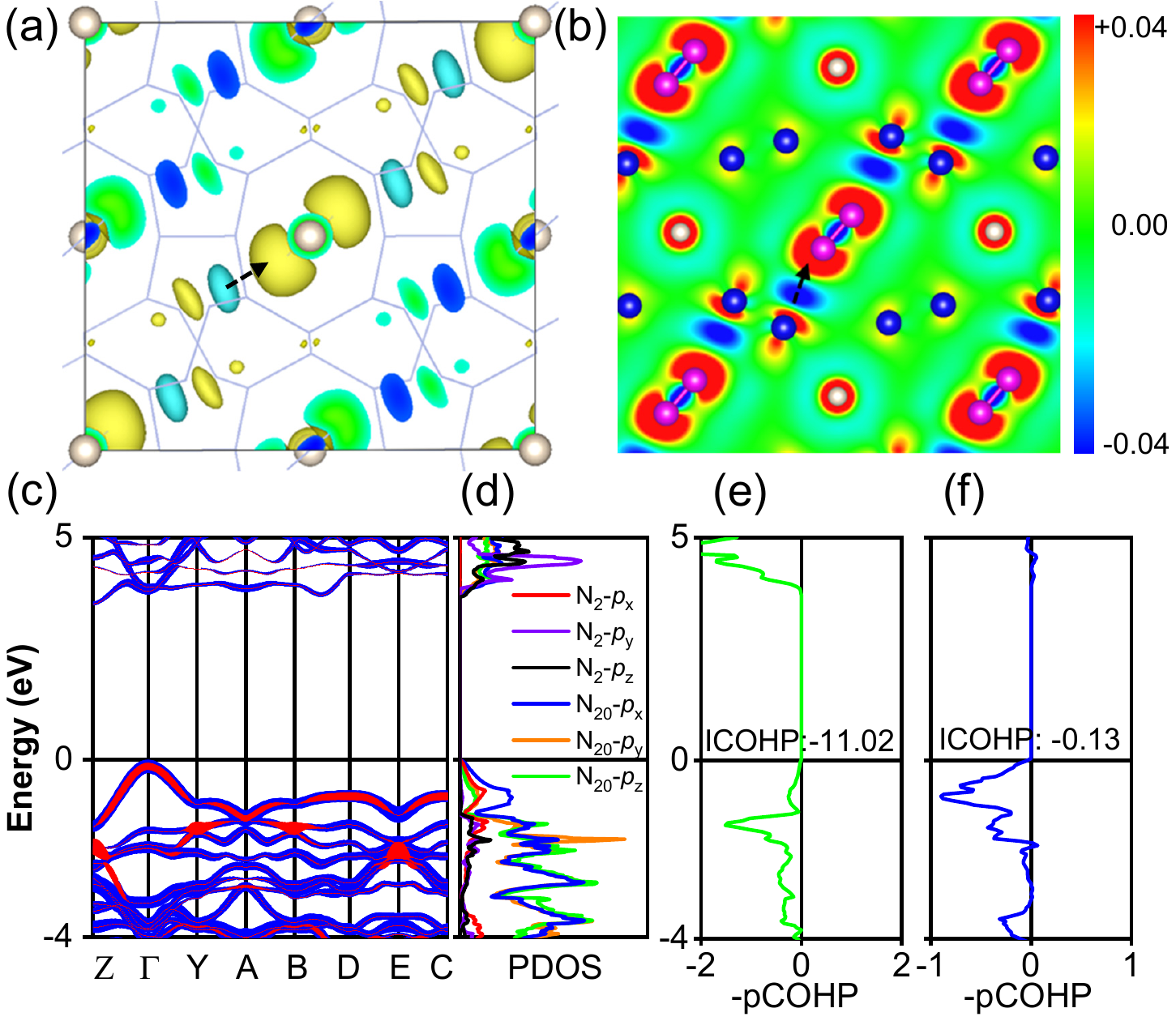}
\caption{\label{Figure3}
 Electronic properties of $C2/m$ HeN$_{22}$ at 100 GPa. (a), (b) As the dotted black arrow indicted, the charge density difference distribution shows that a notable charge transfers from N$_{20}$ cages to N$_{2}$ dimers. (c), (d) Band structure and the projected density of states (PDOS). The orientation of N$_{2}$ dimers is set as $x$ axis. (e), (f) The projected crystal orbital Hamiltonian population (pCOHP) for the shortest inter-cage N3-N4 bonds in (e) and those (N1-N2) between N$_{2}$ dimers and N$_{20}$ cages in (f).}
\end{center}
\end{figure}

The additional evidence for charge transfer in $C2/m$ HeN$_{22}$ is revealed by charge density difference distribution defined as $\Delta$$\rho$ = $\rho$(HeN$_{22}$)$-$$\rho$(He)$-$$\rho$(N$_{2}$)$-$$\rho$(N$_{20}$), where $\rho$(HeN$_{22}$), $\rho$(He), $\rho$(N$_{2}$), and $\rho$(N$_{20}$) are the total charge densities of the whole structure, the sublattices of He, N$_{2}$ dimers, and polymeric N$_{20}$ cages. Charge transfer can be examined from the region of negative $\Delta$$\rho$ to that of positive $\Delta$$\rho$ \cite{R45}. As shown in Fig.~\ref{Figure3}(a) and~\ref{Figure3}(b), the N$_{2}$ dimers were surrounded with the big bubble-like charge accumulations showing a significant positive differential charge (colored in yellow), while the small ellipsoid-shaped differential charge was attached to the vertices of N$_{20}$ cages [Fig.~\ref{Figure3}(a)]. Analyzing regions of charge accumulation and regions of charge depletion, we see that overall charge is transferred from N$_{20}$ cages (lone pairs dominantly) to N$_{2}$ dimers. This result is consistent with Bader charge analysis.

The orbital-resolved band structure shows that $C2/m$ HeN$_{22}$ is an insulator with DFT band gap of 3.67 eV at 100 GPa [Fig.~\ref{Figure3}(c)]. The highest valence band is due to hybridization of $2p$ states between N$_{2}$ dimers and N$_{20}$ cages [Fig.~\ref{Figure3}(d)], implying the existence of fundamental interaction. To determine such interaction, the COHP curves were calculated for bonding character analysis. A negative COHP indicates bonding while positive one indicates antibonding \cite{R46}. The calculated projected COHP curves (pCOHP) for the shortest N-N of N$_{20}$ cages, or between N$_{2}$ dimers and N$_{20}$ cages are both positive in the vicinity of the Fermi level ($E_{F}$), clearly indicative of antibonding states [Fig.~\ref{Figure3}(e) and~\ref{Figure3}(f)]. The integrated COHP (ICOHP) up to the $E_{F}$ is $-$11.02 eV/pair for the shortest inter-cage N3-N4 [Fig.~\ref{Figure3}(e) and~\ref{Figure2}(d)] while it is $-$0.13 eV/pair for the nearest N1-N2 between N$_{2}$ dimers and N$_{20}$ cages [Fig.~\ref{Figure3}(f) and~\ref{Figure2}(d)]. This clearly suggests that the inter-cage N-N bonds are strong and covalent, whereas the results are consistent with a weak interaction with some degree of ionicity (charge transfer) between N$_{2}$ dimer and N$_{20}$ cage. Therefore, our notation N$_{20}$$^{\delta+}$N$_{2}$$^{\delta-}$ is confirmed once again.

\begin{figure}[t]
\begin{center}
\includegraphics[width=8.0cm]{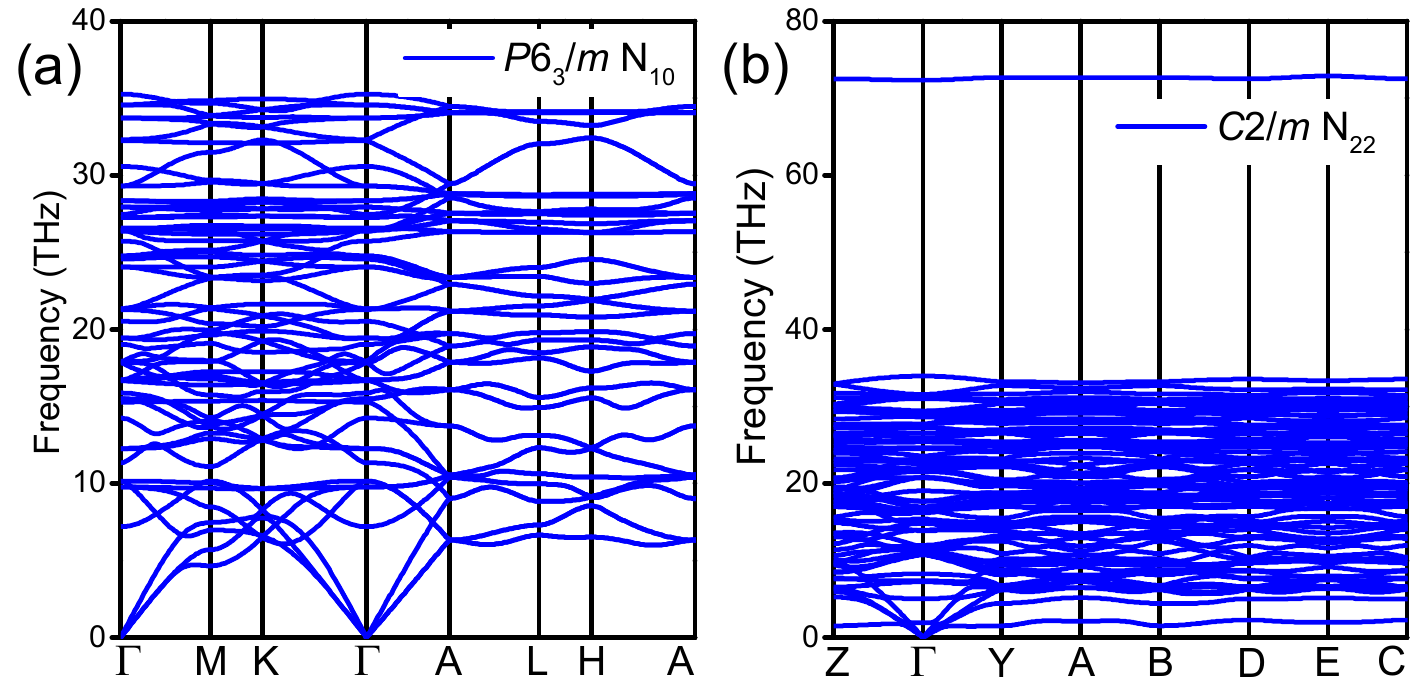}
\caption{\label{Figure4}
Phonon dispersion curves of (a) $P6_{3}/m$ N$_{10}$ and (b) $C2/m$ N$_{22}$ at zero pressure.}
\end{center}
\end{figure}

For all predicted helium-bearing nitrogen-rich compounds, it is natural to consider whether or not these high-pressure phases can be quenchable to ambient pressure, and to examine the kinetic stability of the nitrogen framework after the removal of helium atoms. Therefore, we calculated phonon dispersion curves for these structures at relevant pressure ranges. The results show that $P4/ncc$ HeN$_{4}$ is dynamically stable at 50 GPa because there are no imaginary phonon frequencies in the whole Brillouin zone, and so $R\overline{3}m$ HeN$_{6}$, $P6_{3}/m$ HeN$_{10}$, and $C2/m$ HeN$_{22}$ at 100 GPa (Fig.~S4) \cite{R35}. Interestingly, phonon dispersion curves show that polymeric HeN$_{10}$ and HeN$_{22}$ are dynamically stable at zero pressure (Fig.~S5) \cite{R35}, whereas $P4/ncc$ HeN$_{4}$ and $R\overline{3}m$ HeN$_{6}$ phases are not (and thus zero-pressure properties of these unstable phases will not be discussed further). Moreover, the relaxed pure nitrogen structures, $P6_{3}/m$ N$_{10}$ and $C2/m$ N$_{22}$ are both dynamically stable at ambient pressure (Fig.~\ref{Figure4}). To further examine the thermal stability of $C2/m$ HeN$_{22}$, \textit{ab initio} molecular dynamics simulations show that (HeN$_{20}$)$^{\delta+}$N$_{2}$$^{\delta-}$ structure is thermally stable up to 1000 $K$ at ambient pressure (Fig.~S6) \cite{R35}. The existence of superior thermal and dynamical stability, together with the already synthesized precursor of vdW compound of He(N$_{2}$)$_{11}$, imply that $C2/m$ HeN$_{22}$ or $P6_{3}/m$ HeN$_{10}$ may be quenchable to ambient pressure regardless both with and without the removal of helium. The polymeric ground-state structures of $P6_{3}/m$ N$_{10}$ and $C2/m$ N$_{22}$ are therefore natural candidates for high-energy-density materials. The energy difference between a material and its decomposition products is utilized to evaluate the energy density \cite{R07,R47}. On the basis of such definition, the ground-state structures of HeN$_{10}$ and HeN$_{22}$ have the energy densities of 10.44 kJ/g and 10.72 kJ/g respectively, while their helium-released structures possess the same energy density of 10.44 kJ/g, showing that the partially charged N$_{2}$$^{\delta-}$ dimers don't make a contribution to the energy density owing to the weak interaction with the neighboring N$_{20}$$^{\delta+}$ cages. The estimated energy density of $C2/m$ N$_{22}$ or $P6_{3}/m$ N$_{10}$ is $\sim$2.4 times larger than that of TNT (4.3 kJ/g) or 1.8 times larger than that of 1,3,5,7-Tetranitro-1,3,5,7-tetrazoctane (HMX, 5.7 kJ/g) \cite{R48}. In addition, the detonation velocity ($D$) and pressure ($P$), as other important factors for the performance of high-energy-density materials, are estimated from the empirical Kamlet$йд-$Jacobs equations \cite{R49} with $D$ = 1.01($NM$$^{0.5}$$Q$$^{0.5}$)$^{0.5}$(1+1.30$\rho$) and $P$ = 15.58$\rho$$^{2}$$NM$$^{0.5}$$Q$$^{0.5}$, where $\rho$, $Q$, $N$, and $M$ represent the density, energy density, moles of dinitrogen gas per gram of explosives (mol/g) and the molar mass (g/mol) for dinitrogen gas \cite{R47,R50}, respectively. As shown in Table SIII \cite{R35}, the detonation velocity and pressure for $P6_{3}/m$ N$_{10}$ (21.59 km/s and 2660 kbar) and $C2/m$ N$_{22}$ (22.67 km/s and 3006 kbar) are much higher than those of TNT (6.9 km/s and 190 kbar) and HMX (9.1 km/s and 393 kbar). All these indicate that $P6_{3}/m$ N$_{10}$ and $C2/m$ N$_{22}$ are very promising high-energy-density materials. Finally, given the large activation barrier for synthesis of high-pressure phases of $cg$-N and $HLP$-N \cite{R19,R25,R28}, one can attempt to synthesize these newly exotic structures above 60 GPa and at a temperature higher than 2000 $K$.

In summary, we performed extensive structure searches for the high-pressure phases of helium-nitrogen mixtures and identified four stable compounds of HeN$_{4}$, HeN$_{6}$, HeN$_{10}$, and HeN$_{22}$ at relevant pressure ranges. Among these predicted structures, a polymeric phase of HeN$_{22}$ consists of neutral He, partially charged N$_{2}$$^{\delta-}$ dimers, and N$_{20}$$^{\delta+}$ cages. Particularly, the unique nitrogen clathrate salt (N$_{20}$$^{\delta+}$N$_{2}$$^{\delta-}$) is kinetically stable upon decompression to ambient pressure after removing helium, which provides an unexpected structure induced by helium, deepens the comprehension of nitrogen phase diagram, and proposes a route to the discovery of potential high-energy density materials.

This work was supported by the National Key R\&D Program of China (Grant No.2018YFA0703400), the National Science Foundation of China (Grants 11674176, 11874224, and 21803033), the Tianjin Science Foundation for Distinguished Young Scholars (Grant No. 17JCJQJC44400) and Yong Elite Scientists Sponsorship Program by Tianjin (No. TJSQNTJ-2018-18). A.R.O. thanks Russian Science Foundation (Grant No. 19-72-30043). X.D. and X.F.Z thank the computing resources of Tianhe II and the support of Chinese National Supercomputer Center in Guangzhou.



\end{document}